# Cross-border cannibalization: Spillover effects of wind and solar energy on interconnected European electricity markets

**Clemens Stiewe[a,*], Alice Lixuan Xu[a], Anselm Eicke[a], Lion Hirth[a,b]**

[a]*Centre for Sustainability, Hertie School, Friedrichstraße 180, 10117 Berlin, Germany*
[b]*Neon Neue Energieökonomik GmbH, Karl-Marx-Platz 12, 12043 Berlin, Germany*

*Corresponding author: stiewe@hertie-school.org

The average revenue, or market value, of wind and solar energy tends to fall with increasing market shares, as is now evident across European electricity markets. At the same time, these markets have become more interconnected. In this paper, we empirically study the multiple cross-border effects on the value of renewable energy: on one hand, interconnection is a flexibility resource that allows to export energy when it is locally abundant, benefitting renewables. On the other hand, wind and solar radiation are correlated across space, so neighboring supply adds to the local one to depress domestic prices. We estimate both effects, using spatial panel regression on electricity market data from 2015 to 2023 from 30 European bidding zones. We find that domestic wind and solar value is not only depressed by domestic, but also by neighboring renewables expansion. The better interconnected a market is, the smaller the effect of domestic but the larger the effect of neighboring renewables. While wind value is stabilized by interconnection, solar value is not. If wind market share increases both at home and in neighboring markets by one percentage point, the value factor of wind energy is reduced by just above 1 percentage points. For solar, this number is almost 4 percentage points.

# 1 Introduction

Wind and solar electricity are produced at practically zero marginal cost, which is a blessing for consumers but can be a curse for investors. As the price of electricity drops on windy and sunny days with abundant renewable generation due to the merit-order effect, the revenue of renewable generators is reduced too. The market value of renewable energy therefore tends to decline at increasing renewable market penetration (Borenstein, 2008; Lamont, 2008). It is hence argued that the market value of renewables could fall faster than costs, so that continued subsidies for renewables would be needed to meet policy targets (Sivaram & Kann, 2016; Chyong et al., 2019).

The European Union (EU) has seen a remarkable rise in renewables deployment recently, as wind and solar have increased their combined market share from 11% in 2015 to 23% in 2022 (Eurostat, 2024). Meanwhile, national European electricity markets have become increasingly interconnected with the buildout of cross-border interconnectors and the operational coupling of electricity spot markets through advanced clearing algorithms. Both the deployment of renewables and the creation of an integrated electricity market are at the heart of EU energy policy (Zachmann et al., 2024).

In this paper, we estimate the effects that the simultaneous surge in renewable market penetration and interconnector capacity has had on the average market revenue, or market value, of wind and solar generators. Theory suggests that the effects are two-fold: On the one hand, well-connected markets can export surplus renewable electricity more easily, which mitigates the drop of market value at increasing market penetration. On the other hand, higher connectedness also increases the price pressure from neighboring renewables as wind and solar are correlated across space, which exacerbates the value drop. To isolate cross-border effects, we control for other features of electricity systems that moderate the effect of renewable market penetration on its market value, such as flexible hydroelectricity.

We contribute to the literature by being the first to empirically model the spatial dimension of renewable cannibalization. Additionally, we measure the spatial effects across a rich panel of 30 interconnected European electricity markets, using data from 2015 to 2023. A third novelty of this study is that we provide a first empirical estimate of the moderating effect of market connectedness on domestic and cross-border cannibalization effects.

We find significant domestic and cross-border effects of onshore wind electricity (hereinafter referred to as wind) generation on wind market value across Europe. A one percentage point (p.p.) increase in domestic wind market penetration causes a 0.62 p.p. drop in wind value factors, which measure the market value of wind relative to average electricity prices. In addition to that, a one p.p. increase in wind market penetration across neighboring markets causes a 0.48 p.p. drop in domestic wind value. For solar, both effects are larger in size, 1.39 and 2.38, respectively. In line with theory, connectedness has a two-fold effect, mitigating the negative effect of domestic market penetration but exacerbating the effect of neighboring market penetration on domestic market value. While the mitigating and exacerbating effects are of equal size for solar, the mitigating effect is slightly stronger for wind, signaling gains for wind from higher market connectedness.

The remainder of this paper is structured as follows: Section 2 reviews the literature on the value drop of renewable energy and its moderators. Section 3 introduces the model variables and their expected effects, lays out our identification strategy, describes the data used, and explains our choice of an econometric model. Section 4 presents and discusses our results. Section 5 concludes.

# 2 Literature

Weather-dependent and intermittent wind and solar energy are generated at virtually zero marginal cost. Generators with high fuel costs are therefore replaced by wind and solar energy on a sunny and windy day, reducing the spot price of electricity (Sensfuß et al., 2008; Woo et al., 2011). This also means lower market revenue for wind and solar generators on that day. A large body of literature has investigated the negative effect of renewable market penetration on renewable market value, often referred to as "cannibalization effect". Earlier reviews of the topic can be found in Borenstein (2008), Lamont (2008), Nicolosi (2012), and Hirth (2013). Subsequent empirical studies mostly use time-series econometric methods and find a drop in both wind and solar market value at increasing market penetration in various markets such as Germany (Zipp, 2017), Italy (Clò & D'Adamo, 2015), Spain (Peña et al., 2022), and California (López Prol et al., 2020). The cannibalization of solar value is usually stronger than the cannibalization of wind value, as wind energy is generated more evenly throughout the day while solar generation concentrates around noon (López Prol & Schill, 2021).

While much research finds that the penetration rate is an important driver of wind and solar market value, studies have also identified numerous additional factors that drive market value and/or moderate the value drop. Flexible energy storage and consumption that reacts to scarce or abundant renewable electricity supply can substantially mitigate the value drop. Energy storage in general (Korpaas et al., 2003) and in particular hydro (Benitez et al., 2008; Hirth, 2016; Schöniger & Morawetz, 2022) and thermal (Mills & Wiser, 2013) storage explain why some markets manage to integrate wind and solar energy better than others. The positive effect of flexible electricity consumption on renewable market value is well-established too (Garnier & Madlener, 2016; Hirth & Radebach, 2016; Ruhnau, 2022).

Renewable market value is also driven by fuel costs for non-renewable generators. A steep electricity supply curve elevates prices and thus inframarginal revenues for renewables but exacerbates the value drop with increasing renewable penetration. Supply curves in electricity are usually steep when gas prices are high, coal prices are low, and carbon prices are low (Hirth, 2013). Higher carbon prices therefore mitigate the value drop of wind and solar (Liebensteiner & Naumann, 2022).

A smoother wind generation profile and hence a more stable wind market can be achieved if wind plants are spread across larger areas because wind speeds are locally different (Katzenstein et al., 2010). Advanced turbine design can also mitigate the intermittency of wind generation (Hirth & Müller, 2016). There is no meaningful geographic smoothing for solar generation, except for a slightly higher coefficient of variation at higher latitudes, as a result of higher seasonal solar variation (López Prol et al., 2024).

Wind and solar also affect each other's market value. López Prol et al. (2020) find that wind penetration reduces solar market value in California, while solar penetration increases wind market value. Similarly, Schöniger and Morawetz (2022) find that in Germany, Great Britain and Greece, solar infeed increases prices when wind infeed is high. Wind penetration reduces the market value of Spanish solar and vice versa (Peña et al., 2022).

In addition to the price effects of domestic renewable generation, a body of literature has also identified cross-border price effects. Frauendorfer et al. (2018) find price-depressing effects of renewable electricity imports from Germany on Swiss electricity prices. The same effect has also been observed in Austria (Würzburg et al., 2013) and France (Pham, 2019). However, these studies do not model the effects on renewable market value. Additionally, most of them focus on a pair of neighboring markets, with the exception of Keles et al. (2020), who examine all three neighbors of the Swiss electricity market and find negative cross-border effects on Swiss prices by renewable in-feed in the German-Austrian-Luxemburgish and French markets.

On the other hand, a price-lifting effect is found for markets that export renewable electricity. A modelling study of interconnected electricity markets by Tveten et al. (2016) finds that the strong interconnection between wind-rich Denmark and its neighbors explains its high wind market value. This confirms Green and Vasilakos (2012), who show numerically that Denmark is effectively using its neighbors for short-term storage to offset the intermittency in its wind output, which becomes especially salient when Danish prices drop more during hours of congested transmission lines to the north of Denmark. Their findings are supported by Schöniger and Morawetz (2022), who also attribute stable Danish prices to high export capacity. The size of those effects should be higher for markets with higher import/export capacity. Indeed, modelling studies find that increasing interconnection between imperfectly correlated markets lifts wind market value in the exporting market (Obersteiner, 2012), and that expanding interconnection stabilizes wind but not solar value (Hirth, 2015).

The cross-border effects of renewable generation have been estimated extensively in time-series analyses, such as in Pham (2019) and Keles et al. (2020). This complicates controlling for important system characteristics such as hydro flexibility, which hardly vary over time. Schöniger and Morawetz (2022) do capture these characteristics in their panel analysis, both by including them in a pooled model and by ex-post comparison of fixed effects (FE) estimates with country-level hydro shares and other moderators. Random effects models of the Mundlak (1978) type, which allow to jointly obtain FE-like estimates for time-varying variables as well as estimates for time-invariant variables, are another, not yet applied, way to tackle this econometric challenge.

In this paper, we provide the first joint empirical estimate of domestic and cross-border effects of renewable market penetration on its market value. Moreover, we measure these effects across a panel of interconnected European markets. This allows us to causally estimate how different levels of connectedness between those markets moderate the domestic and cross-border effects of renewable market penetration.

# 3 Data and Method

In the following section, we introduce the model variables and respective hypotheses, delineate our identification strategy, describe the data used, and explain our choice of econometric model.

## *3.1 Model variables and expected effects*

We use an econometric panel model to estimate the domestic and spatial effects of renewable market penetration on renewable market value across Europe. Interactions of domestic and neighboring market penetration with interconnector capacity capture the moderating effect of connectedness. Control variables include flexibility from pumped storage and reservoir hydro, the variability of renewable generation, the correlation between renewable generation and electricity consumption, as well as the ratio between carbon-adjusted gas and coal prices to proxy the slope of the electricity supply curve. A separate model is estimated for wind and solar value. In the following we explain the structure of our panel as well as our model variables and their expected effects.

Electricity bidding zones represent the panel entities in our analysis. Electricity spot markets in Europe are organized in a zonal system, where all electricity producers and consumers within the geographic area of a so-called bidding zone can trade electricity at the spot market without grid constraints. While most European bidding zones reflect national borders, Norway, Sweden, Denmark and Italy have multiple bidding zones. Germany and Luxemburg currently share a unified bidding zone, which included Austria until the split of the German-Austrian-Luxemburgish zone in 2018. We include each of these bidding zones as a separate panel entity in our analysis, making it an unbalanced panel.

Market value (also called “capture price”) is a metric used to approximate the income earned by electricity generators from selling their output on the wholesale spot market, without additional subsidies (Joskow, 2011). It reflects the possible market revenue from selling one unit of generated energy (usually: one megawatt hour), aggregated over a certain period. The choice of aggregation period for market value substantially alters the estimated effects of market penetration and is hence not an innocent one. Aggregation periods are different across the literature, from daily (Liebensteiner & Naumann, 2022) to monthly (Winkler et al., 2016) or annual (Hirth, 2013). The longer the aggregation period, the better one approximates an investor's perspective, who is interested in earnings over the power plant's lifetime. Annual and higher aggregates leave little variation to exploit empirically, however. Daily aggregates, on the other hand, offer substantial variation but underestimate the effects of market penetration on market value relative to average prices: At higher temporal resolution, renewable generation increasingly affects not only its own market value, but also average prices. Particularly for wind generation, which is usually spread out across most of the day, one finds a smaller effect on relative revenue. We hence aggregate market value at a monthly level and discuss the sensitivity of our results to daily and annual aggregation in 4.3.3.

Monthly market value $MV_{i,t}^{\{w,s\}}$ per generation technology wind/solar $\{w, s\}$ and bidding zone $i$ is defined as the generation-weighted spot price, i.e., the monthly sum of the product of hourly wind/solar generation $g_h^{\{w,s\}}$ and contemporaneous spot prices $p_h$, divided by the monthly sum of hourly wind/solar generation:

$$MV_{i,t}^{\{w,s\}} = \frac{\sum_{h=1}^{H} p_h * g_h^{\{w,s\}}}{\sum_{h=1}^{H} g_h^{\{w,s\}}}$$

Where $H$ is the number of hours in a month. In this study, we measure market value as the relative price of wind/solar electricity compared to average electricity prices. Using this "value factor" (also called "capture rate") as the dependent variable in our model corrects for unobserved shocks to average price levels following business cycles and events like the 2022/2023 European energy crisis. The value factor $VF_{i,t}^{\{w,s\}}$ is defined as the ratio of market value and average monthly spot prices:

$$VF_{i,t}^{\{w,s\}} = \frac{MV_{i,t}^{\{w,s\}}}{\bar{p}} = \frac{MV_{i,t}^{\{w,s\}}}{\sum_{h=1}^{H} p_h/H}$$

Domestic market penetration is defined as the monthly sum of hourly wind/solar generation $g_h^{\{w,s\}}$ divided by the monthly sum of hourly electricity consumption $c_h$:

$$P_{D}{}_{i,t}^{\{w,s\}} = \frac{\sum_{h=1}^{H} g_h^{\{w,s\}}}{\sum_{h=1}^{H} c_h}$$

The generation data we use therefore includes all instances of economic and grid-related curtailment. In case of significant curtailment, actual generation shares will thus be larger than potential generation shares. The implications of that are discussed in 3.3.

We measure connectedness of electricity markets to neighboring markets by approximating total interconnector capacity of a bidding zone. Across most of Europe, electricity trade between bidding zones is organized through flow-based market coupling[1]. This is an algorithm that minimizes operational costs with respect to trade flows subject to a large number of technical constraints (including, but not limited to limits on interconnectors) (Schönheit et al., 2021). As a consequence, interconnector capacity between two zones is not a fixed technical parameter. Hence, we approximate interconnector capacity from the largest observed trade flows ("scheduled commercial exchanges") during the individual hours of a year. More precisely, we define our interconnector capacity proxy as the 95% quantile of hourly absolute net exports from bidding zone $i$ to $j$ during the first sample year[2], summed over all its direct neighbors. To reflect that a 1 GW interconnector is more relevant for a smaller than for a larger bidding zone, we normalize approximated interconnector capacities with mean annual zonal load. European bidding zones and their approximated interconnector capacity are shown in Figure 1.

[1] In 2015, flow-based market coupling was first introduced in Belgium, the Netherlands, France, Germany, and Austria in 2015 and expanded to Croatia, the Czech Republic, Hungary, Luxembourg, Poland, Romania, Slovakia and Slovenia in 2022.

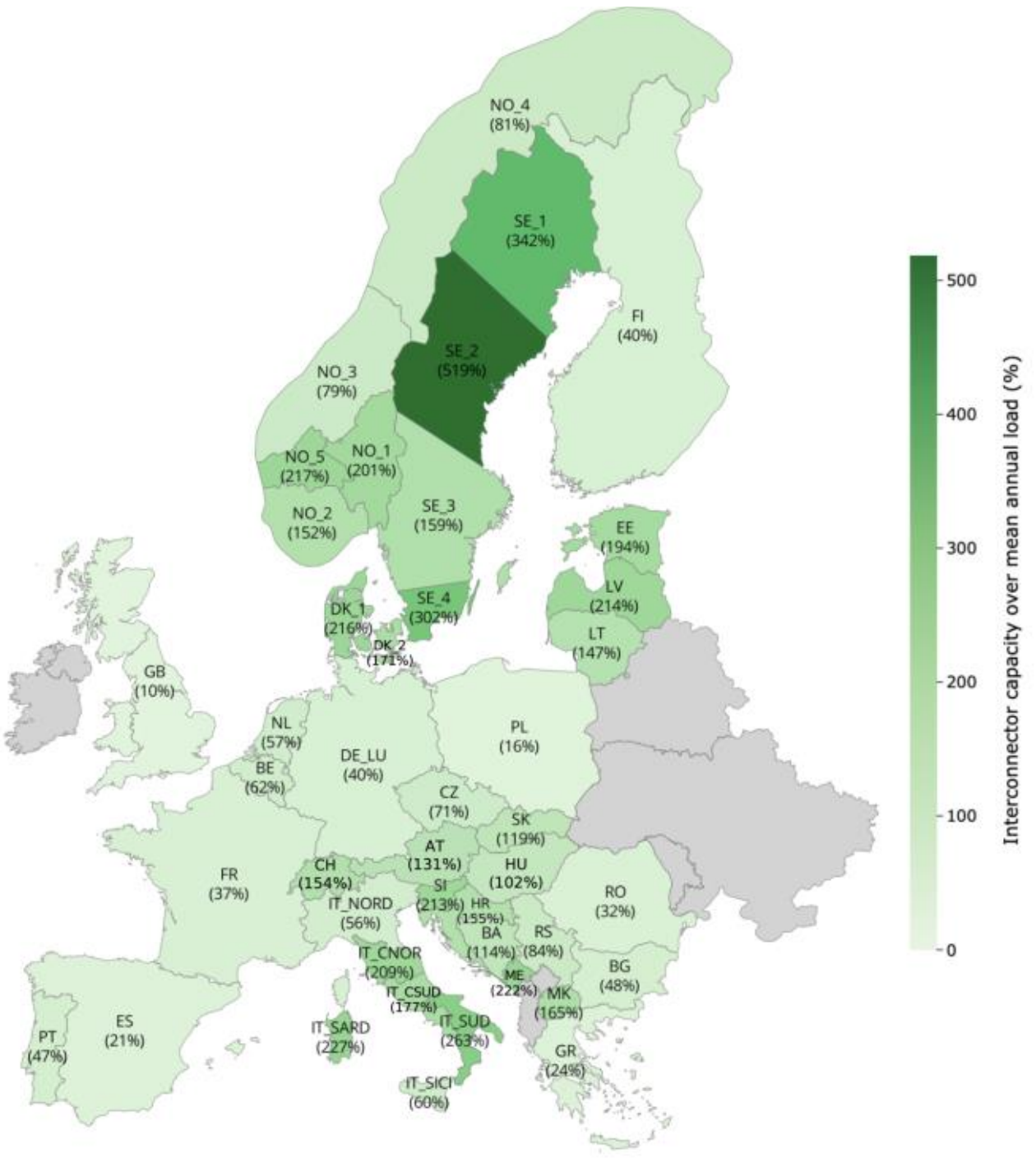

*Figure 1: European bidding zones and interconnector capacity*

We expect higher connectedness to reduce the negative effects of domestic wind market penetration on domestic wind value: A well-interconnected zone can better export excess wind electricity in windy hours to increase the market value of wind, while higher imports of wind electricity from neighboring zones when there is little domestic wind generation reduce average prices. Both exports in windy hours as well as imports in windless hours increase the value factor of wind, which is defined as the ratio of market value and average prices. At the same time, we expect higher connectedness to increase the negative effects of neighboring wind penetration on domestic wind value: Other things equal, a well-interconnected zone will see higher wind electricity imports when domestic wind generation is high, pushing down domestic wind value further. Higher exports in hours with little domestic wind generation increase average prices, which reduces the wind value factor. We capture this two-fold effect through two interaction terms: Interconnector capacity with domestic wind/solar market penetration and interconnector capacity with neighboring wind/solar market penetration. In light of the above, we except the first interaction to have a positive sign and the second to have a negative sign.

As controls we include all variables that previous authors have identified as factors shaping the market value of wind and solar energy. Both their main effects on value factor levels as well as their interactions with domestic wind (solar) penetration – to capture the moderating effect on domestic cannibalization – are included in our wind (solar) model.

---

[2] The first sample year is 2015 for all bidding zones except the German-Luxemburgish (DE_LU) and Austrian (AT) bidding zones, which were created in 2018 after the split of the German-Austrian-Luxemburgish zone (DE_AT_LU).

We control for the installed capacity of pumped storage and reservoir hydro power plants at the bidding zone level to capture their effect on the wind/solar value drop, which we expect to be positive. Like interconnector capacity, we also normalize installed hydro capacity by mean annual zonal load.

To capture the potentially mitigating effect of higher electricity demand on the value drop, we correlate the vectors of hourly wind/solar generation and hourly electricity consumption to compute a monthly average coefficient of correlation and include it as another control.

The monthly coefficient of variation of hourly wind/solar generation is used as a proxy for the variability of wind/solar generation on an hourly basis. We expect a higher coefficient of variation (i.e., a more "peaky" generation pattern) to exacerbate the value drop of wind and solar.

Cross-technology effects of solar market penetration on wind value – and vice versa – are modelled by including both domestic and neighboring wind (solar) penetration in our solar (wind) models. Following the literature, we expect a positive effect of solar penetration on wind value and a negative effect of wind penetration on solar value.

We approximate the slope of the supply curve with the ratio of Dutch TTF natural gas and coal API2 prices. We add the EU ETS price, multiplied by respective carbon contents, to both coal and gas prices. We hence assume that all European bidding zones face the same clean gas and coal prices. A higher clean gas-coal price ratio results from higher gas, lower coal, or lower carbon prices and means a steeper supply curve, which is expected to exacerbate the drop of wind/solar value with increasing market penetration.

## *3.2 Modelling spatial effects*

In addition to the domestic effects of renewable market penetration on value factor, we expect spillover effects of renewable generation between neighboring and interconnected European electricity markets. In other words, we want to assess if and to what extent wind and solar generation in neighboring bidding zones adds to the downwards pressure on domestic value factors that comes from domestic renewable generation. More specifically, we are interested in the spatial effects of an independent variable, wind/solar market penetration, on value factor. This can be achieved with a *spatial lag of X* (SLX) model (Anselin et al., 2008; Elhorst, 2014; Beenstock & Felsenstein, 2019). The SLX model captures local spillover effects between panel entities by including the values of independent variables in geographically adjacent panel units.

In our wind and our solar models, we therefore regress value factor both on domestic and neighboring (i.e., "spatially lagged") wind/solar penetration. The construction of the spatially lagged variables merits some explanation. As a first step we only consider the direct geographical neighbors of a bidding zone, which amounts to a binary spatial weights matrix (Halleck Vega & Elhorst, 2015). However, a simple unweighted average of wind/solar penetration across all direct neighbors does not reflect that bilateral interconnector capacities vary greatly (see Figure 1), which moderates cross-border price effects. We expect smaller interconnector capacity between focal zone $i$ and neighbor $j$ to increase the likelihood of interconnector congestion. During hours of interconnector congestion between $i$ and $j$, additional renewable generation in $j$ will have little effect on prices in $i$ (and vice versa). In fact, well-interconnected zones tend to have wind-rich and solar-poor neighbors in our sample. Using an unweighted average of wind/solar penetration across the border would therefore overestimate the spillover effects of wind and underestimate the effects of solar energy across our sample, which we discuss in 4.3.1.

We therefore weight wind/solar market penetration of $i$'s direct neighbor $j$ ($P_{D}{}_{j,t}^{\{w,s\}}$) by approximated interconnector capacity between $i$ and $j$ ($IC_{ij}$), normalized by total domestic interconnector capacity of zone $i$ ($\sum_{j=1}^{J} IC_{ij}$). Summing up interconnector-weighted and normalized market penetration over all neighbors $J$ gives us the spatially lagged (hereinafter referred to as neighboring) market penetration variable:

$$P_{N}{}_{i,t}^{\{w,s\}} = \sum_{j=1}^{J} (P_{D}{}_{j,t}^{\{w,s\}} * \frac{IC_{ij}}{\sum_{j=1}^{J} IC_{ij}})$$

A one p.p. increase in this variable can thus be interpreted as a one p.p. increase in wind/solar penetration across all of a bidding zone's direct neighbors (i.e., all neighbors simultaneously exhibit an increase in wind/solar penetration of one p.p.), irrespective of the number of neighbors, the absolute level of connectedness of a bidding zone, and the relative differences in bilateral connectedness of a bidding zone across all its neighbors.

While the heterogeneity of wind speed patterns across Europe allows to easily isolate the effects of domestic and neighboring wind, solar output profiles do not vary much across Europe (López Prol et al., 2024). Domestic and neighboring solar penetration are therefore strongly correlated, which likely inflates standard errors[3]. We regard the higher likelihood of erroneously retaining a false null hypotheses as fair price for the additional insights gained from a spatial solar model.

### 3.3 *Identification strategy*

Causal identification can only be achieved if the independent variables in our model are exogenous to electricity prices and thus wind/solar market value. This generally holds for the generation of renewable energy, which depends on weather conditions. When generators voluntarily curtail their output during periods of zero and negative prices, however, the exogeneity assumption is violated. Market-based curtailment figures are not available for Europe but the instrumental variable approach of Liebensteiner & Naumann (2022) suggests no endogeneity bias for Germany induced by curtailment. We thus estimate the effect of actual generation.

Interconnector capacity is not affected by wholesale price fluctuations in the short run. Increasing price differences between neighboring zones, however, e.g. because of renewable energy deployment, signal gains from trade and incentivize the buildout of interconnector capacity in the long run. To address this potential source of endogeneity, we use interconnector capacity in the first respective sample year for each bidding zone in our sample as a time-invariant proxy for connectedness. We follow a similar approach for flexible pumped storage and reservoir hydro plants. Their generation patterns are endogenous to electricity spot prices and hence renewable market value, which is why we use annual installed capacities instead of actual electricity generation.

### 3.4 *Data*

We use electricity market data from 2015 until 2023 provided by the European Network of Transmission System Operators for Electricity (ENTSO-E) Transparency Platform (ENTSO-E, 2024). For Sweden and Spain, data missing on this platform was retrieved from national authorities. After accounting for missing values, 30 different bidding zones remained in our sample. Figure A-1 shows monthly domestic and neighboring wind and solar market penetration for all bidding zones included our wind and solar models. Since not all zones have had significant wind and solar generation during the sample period, 29 bidding zones were included in the wind and 23 in the solar model. The data comprises hourly electricity consumption and day-ahead electricity prices, onshore wind and solar generation, installed capacities of pumped storage and reservoir hydro, and scheduled commercial exchanges of electricity between bidding zones. We test our data for unit roots using the Maddala and Wu (1999) panel unit root test and find that all variables are stationary. Table 1 shows summary statistics of all variables included in our regression models except for neighboring penetration, whose summary statistics mirror domestic penetration. The value drop of wind and solar across Europe during our sample period is clearly visible in Figure 2, which plots wind/solar value factors against monthly domestic wind/solar penetration during 2015-2023, including a linear trend line for each bidding zone.

[3] The correlation of hourly (monthly) domestic and neighboring solar penetration per bidding zone is 0.88 (0.90). For onshore wind it is 0.60 (0.75) on an hourly (monthly) level.

| | **Mean** | **Median** | **Min** | **Max** |
|---|---|---|---|---|
| DEPENDENT VARIABLES | | | | |
| Wind value factor | 0.92 | 0.94 | 0.25 | 1.17 |
| Solar value factor | 1.02 | 1.02 | 0.49 | 2.31 |
| INDEPENDENT VARIABLES | | | | |
| Domestic wind penetration (%)[1] | 13 | 9 | 0 | 154 |
| Domestic solar penetration (%)[1] | 3 | 1 | 0 | 26 |
| Interconnector capacity (%)[2] | 131 | 102 | 10 | 519 |
| Hydro pumped storage capacity (%)[2] | 13 | 7 | 0 | 96 |
| Hydro reservoir capacity (%)[2] | 52 | 16 | 0 | 389 |
| Wind-load correlation | -0.04 | -0.04 | -0.63 | 0.49 |
| Solar coefficient of variation | 0.37 | 0.36 | -0.29 | 0.79 |
| Solar-load correlation | 0.75 | 0.73 | 0.26 | 1.62 |
| Wind coefficient of variation | 1.54 | 1.40 | 0.80 | 4.80 |
| Clean gas-coal price ratio | 1.00 | 0.97 | 0.43 | 2.11 |

[1]Actual generation over total electricity consumption, [2]Capacity over mean annual load

*Table 1: Summary statistics*

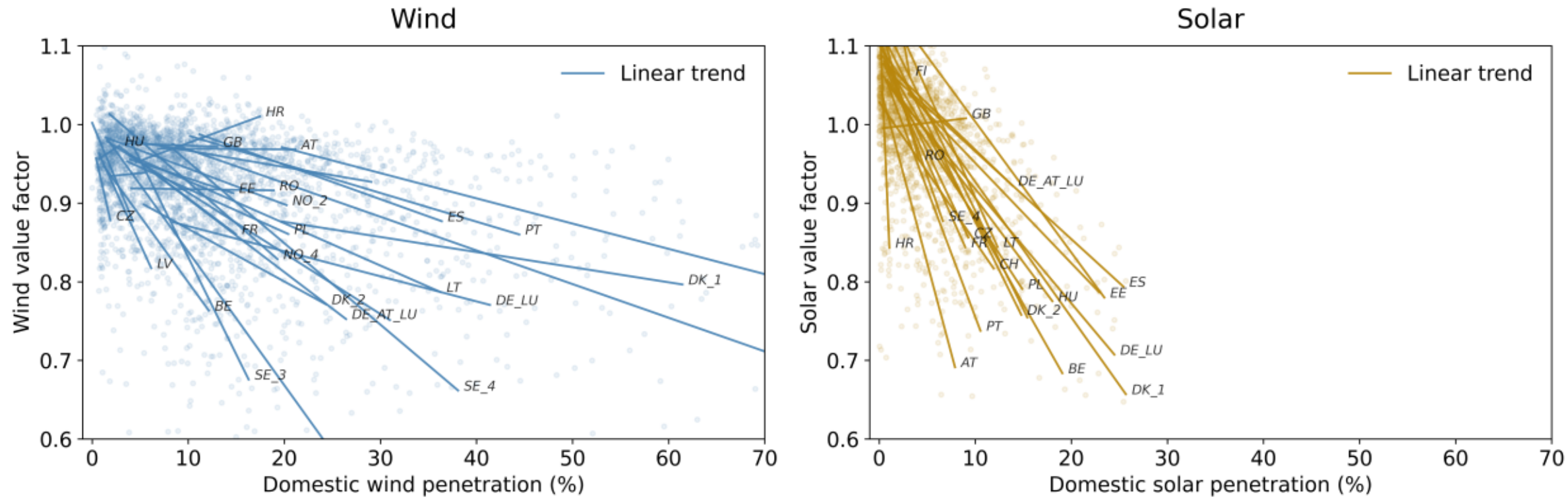

*Figure 2: Monthly wind/solar value factor vs. market penetration for European bidding zones*

## 3.5 Econometric model

Fixed effects models are the tool of choice for most researchers to estimate unbiased effects across panels of heterogenous entities. In our setting, however, this common approach is not ideal because relevant variables, such as interconnector or hydro power capacities, only vary little over time. Most variation is found in average levels between zones (e.g. Norwegian zones have much higher relative interconnector and hydro capacity than France or Germany). Including zone FE would eliminate any higher-level (i.e., *between* zones) variation. We therefore estimate a random effects within-between (REWB) model, which allows to exploit higher-level variation and to obtain FE-like estimates for variables that vary over time (Mundlak, 1978; Bell & Jones, 2015; Wooldridge, 2019). The REWB model separately estimates the effects of variation in independent variable $X_{i,t}$ over time *within* panel entities and *between* panel entities. The *within* effect is obtained by centering $X_{i,t}$ around its entity time average $\bar{X}_i$ and performing regression on the transformed $\ddot{X}_{i,t} = X_{i,t} - \bar{X}_i$. This resembles the *within* transformation in a FE model. Hence, the estimated effect of $\ddot{X}_{i,t}$ is identical to the one obtained by FE estimation and controls for the effects of unobserved heterogeneity at the zone level (Jordan & Philips, 2023). In 4.4 we estimate our model (without time-invariant variables) using FE and show that this holds. *Between* effects in the REWB model are obtained by including the entity means $\bar{X}_i$.[4] Other time-invariant regressors, such as interconnector capacity, can be modelled as well.

We estimate the following REWB model for wind and solar value factors:

$$\begin{aligned} VF_{i,t}^{\{w,s\}} &= \beta_0 + \beta_1 \ddot{P}_{D\,i,t}^{\{w,s\}} + \beta_2 \overline{P_D}_i^{\{w,s\}} + \beta_3 \ddot{P}_{N\,i,t}^{\{w,s\}} + \beta_4 \overline{P_N}_i^{\{w,s\}} \\ &\quad + \beta_5 IC_i + \beta_6 \ddot{P}_{D\,i,t}^{\{w,s\}} * IC_i + \beta_7 \ddot{P}_{N\,i,t}^{\{w,s\}} * IC_i \\ &\quad + \beta' C + \beta' \, \ddot{P}_{D\,i,t}^{\{w,s\}} * \ddot{C} + \beta' \ddot{P}_{D\,i,t}^{\{w,s\}} * \bar{C} + \varepsilon_{i,t} \end{aligned}$$

Where $\ddot{P}_{D\,i,t}^{\{w,s\}} = P_{D\,i,t}^{\{w,s\}} - \overline{P_D}_i^{\{w,s\}}$ are the *within* effects of wind/solar penetration for bidding zone $i$ at time $t$. In addition to these *within* effects, bidding zone time averages $\overline{P_D}_i^{\{w,s\}}$ are included to capture the variation in wind/solar penetration *between* bidding zones. Analogously, the estimates of $\ddot{P}_{N\,i,t}^{\{w,s\}}$ and $\overline{P_N}_i^{\{w,s\}}$ capture the *within* and *between* effects of variation in neighboring wind/solar penetration. Two cross-level interactions of domestic and neighboring market penetration with interconnector capacity ($IC_i$) capture the moderating effect of connectedness. $C$ is a vector of controls including hydro reservoir capacity, hydro pumped storage capacity, correlation of load and wind/solar generation, wind/solar coefficient of variation, the clean gas-coal price ratio as well as cross-technology market penetration, i.e. domestic and neighboring wind penetration in the solar model and vice versa. Except for the cross-technology effects of wind and solar, we interact all controls with domestic wind/solar market penetration to capture their moderating effect on the wind/solar value drop. For controls that vary both within zone and between zones, we include interactions of $\ddot{P}_{D\,i,t}^{\{w,s\}}$ both on the lower level (i.e. *within* effect of wind/solar penetration times *within* effect of control) and across levels (i.e. *within* effect of wind/solar penetration times *between* effect of control). This captures both the effect of changes in controls over time (e.g., the buildout of hydro capacity) and the effect of average control levels (e.g., average installed hydro capacity per bidding zone) on the value drop. Following Giesselmann and Schmidt-Catran (2022), we obtain unbiased lower-level interaction effects by first demeaning each variable and then demeaning the product of both variables. We center the *between* variables around their grand mean to be able to interpret the main effects of domestic and neighboring wind/solar penetration as effects at mean levels of independent variables.

For both wind and solar models, panel Durbin-Watson tests indicate serial correlation of residuals. We therefore compute Newey and West (1987) heteroscedasticity and autocorrelation consistent standard errors for both models.

# 4 Results and discussion

In this section we go through our regression results. We first discuss the results of the wind model in detail and then explain our findings for solar. Regression results for variables of interest are shown in Table 2 and a complete list of results is provided in Table 3.

[4] By construction, *within* and *between* variables are uncorrelated. Hence, the *between* effect of average wind penetration does not control for the *within* effect of changes in wind penetration over time.

| | Wind value factor | Solar value factor | |
|---|---|---|---|
| EFFECTS ON WIND VALUE | | | EFFECTS ON SOLAR VALUE |
| Domestic wind | -0.622*** | -1.392*** | Domestic solar |
| Domestic wind*Interconnector capacity | 0.19*** | 0.412** | Domestic solar*Interconnector capacity |
| Neighboring wind | -0.486*** | -2.377*** | Neighboring solar |
| Neighboring wind*Interconnector capacity | -0.133*** | -0.415* | Neighboring solar*Interconnector capacity |

*Table 2: Regression results for variables of interest. Random effects between-within regression model for wind and solar. Significance levels are reported as *** for p < 1%, ** for p < 5%, and * for p < 10%.*

## 4.1 Wind

We find sizable and significant cannibalization of wind value across European markets. For a one p.p. increase of domestic wind penetration over time (i.e., the *within* effect of wind penetration), the value factor drops by 0.62 p.p. Sign and size are as expected, and fall within the value drop estimates reported by Hirth (2013) for thermal- (-1.62 p.p.) and hydro-dominated (-0.22 p.p.) European systems. Our estimate is also virtually identical wind value drop estimated for California by López Prol et al. (2020). The *between* effects of domestic and neighboring wind penetration are also negative but insignificant. We hence only display the *within* effects in Figure 3.

There are substantial cross-border spillover effects of wind penetration. A simultaneous one p.p. increase in wind penetration across all direct neighbors is associated with a 0.48 p.p. drop in domestic value factors, so it is somewhat smaller than the domestic effect. The smaller effect of neighboring wind penetration is indicative of the imperfect correlation of wind speeds across space, which results in a smoother profile of (interconnector-capacity weighted) average neighboring generation. The effects of domestic and neighboring penetration add up to a 1.1 p.p. drop in wind value factors per one p.p. increase in domestic and neighboring wind penetration. Importantly, those are the effects of domestic and neighboring wind penetration at average levels of interconnector capacity.

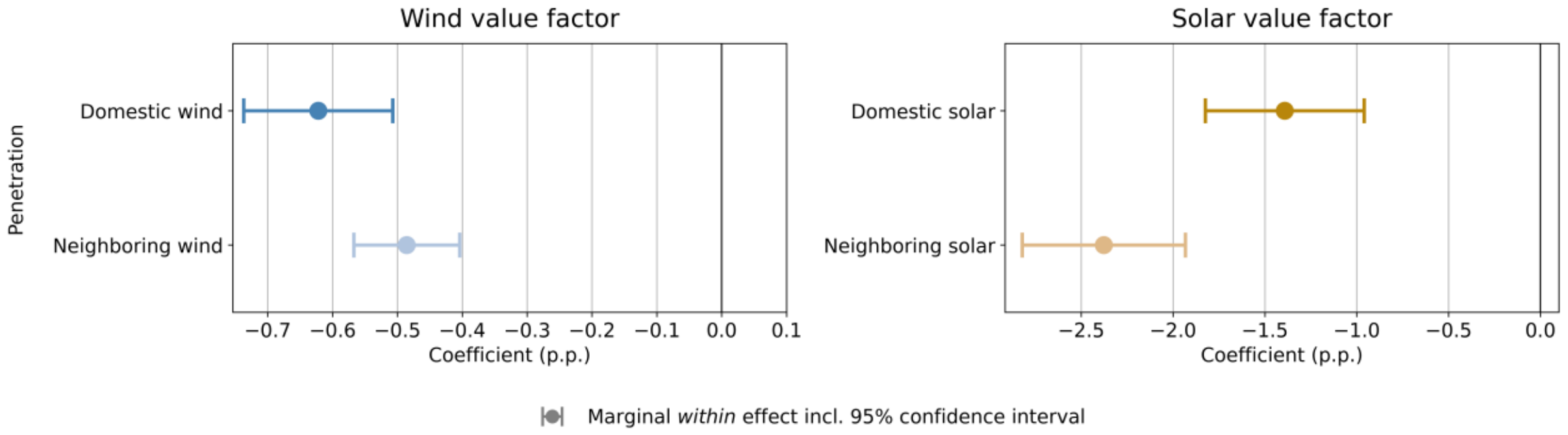

*Figure 3: Marginal effects of domestic and neighboring wind/solar on wind/solar value factors*

The heterogeneity in connectedness across European zones significantly alters the effects of domestic and neighboring wind market penetration. Confirming our expectations, we find that higher interconnector capacity mitigates the domestic value drop but exacerbates the spillovers of neighboring market penetration. The mitigating effect of cross-border trade can be understood by looking at situations of high domestic wind penetration and low neighboring wind penetration (and vice versa): Exports of excess domestic wind electricity in hours with little wind in neighboring zones lift the market value of domestic wind (as additional demand), while wind imports in hours with little domestic wind generation reduce average prices (as additional supply). Both increase the value factor of wind. The exacerbating effect of connectedness on the value drop, on the other hand, takes place in hours where both domestic and neighboring wind penetration are high or low at the same time: Imports of excess wind electricity from neighboring markets push down domestic wind market value further when domestic wind generation is high (as additional supply), while exports in hours with little domestic wind generation hours increase average prices (as additional demand). Both reduce the value factor of wind.

The left panel of Figure 4 shows the conditional effects of domestic and neighboring wind penetration at different levels of normalized interconnector capacity.[5] While the effect of domestic wind penetration on wind value is larger at average interconnector capacity, it becomes smaller with increasing interconnector capacity as both the value-lifting effect in windy hours and the price-depressing effect in windless hours gain in size. The opposite holds for the effect of neighboring wind penetration. The more interconnected a bidding zone, the stronger wind market value is depressed by imports in windy hours. At the same time, exports in windless hours increase average prices and thus reduce the value factor as a measure of normalized revenue. Higher connectedness hence implies a trade-off between the negative effects of domestic and neighboring wind penetration on domestic wind value.

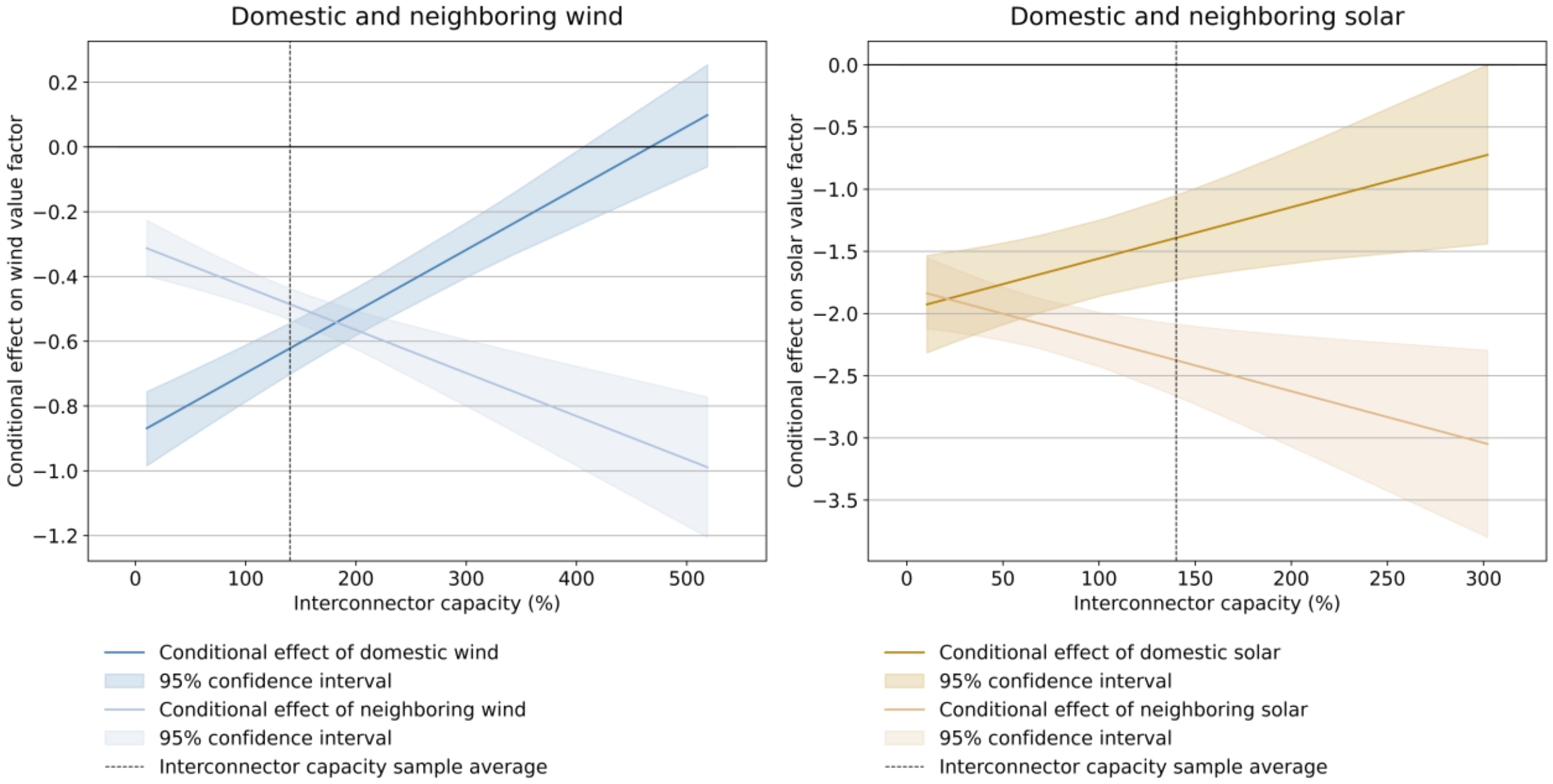

*Figure 4: Effect of domestic and neighboring wind/solar penetration on domestic wind/solar value factors, conditional on interconnection capacity*

This trade-off is not perfect, however: The interaction effect of interconnector capacity with domestic wind penetration is -1.90 and thus larger than the one with neighboring wind penetration at -1.33 (see Table 3). The mitigating effect of connectedness on the wind value drop is hence stronger than the exacerbating effect because wind profiles are imperfectly correlated across space, which makes the aggregate domestic and neighboring wind profile smoother in well-interconnected zones. A smoother profile means a smaller merit-order effect, and hence a less severe value drop for wind energy.

To better understand how the trade-off between domestic and neighboring cannibalization effects materializes at the zone level, we show the conditional effects of domestic and neighboring solar penetration at the respective zonal levels of interconnector capacity in the upper panel of Figure 5. Well-interconnected bidding zones in Denmark and Sweden only see moderate domestic cannibalization because they can export surplus wind generation more easily and experience lower prices in windless hours due to imports. However, their connectedness also makes them more vulnerable against value-depressing imports of wind electricity from neighboring zones when domestic wind generation is high.

[5] The intersections of the dashed line representing the sample average of interconnector capacity with the conditional effect function represent the base effects of domestic/neighboring wind and solar.

Prices zones with less interconnector capacity compared to average electricity consumption, like Poland and the German-Luxemburgish zone, however, exhibit the opposite effect: The constrained ability to export excess wind electricity exacerbates the drop of domestic market value compared to average prices in windy hours and limits the price-reducing effect of imports in windless hours. On the other hand, their below-average connectedness protects domestic wind generators from additional neighboring supply when domestic wind generation is high and limits the potential of exports in windless hours, both of which have a stabilizing effect on domestic wind value. The gains from connectedness for wind are also reflected in the combined point estimates of domestic and neighboring cannibalization: They are around -1.2 p.p. for the least interconnected zones in our sample (like Great Britain, Poland and Spain) and decrease in size to just below -1 p.p. for the most interconnected zones (which are Scandinavian and Baltic zones), although the uncertainty around the latter effect is higher.

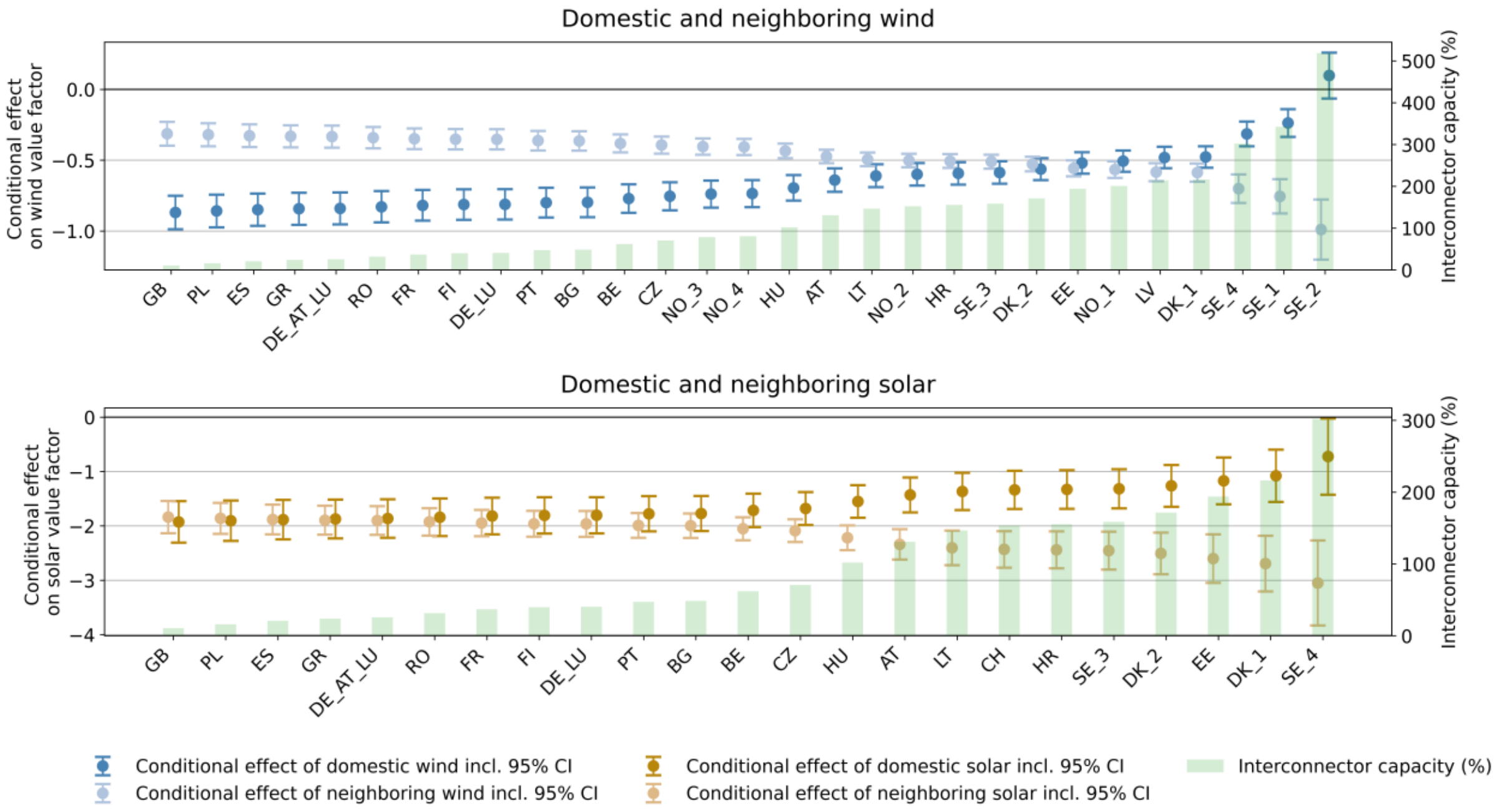

*Figure 5: Zone-level effect of domestic and neighboring wind/solar penetration on domestic wind/solar value factors, conditional on interconnector capacity*

Our control variables help explain why some zones manage to integrate wind better than others. We find that higher average capacity of pumped storage and hydro reservoir plants significantly mitigates the value drop of wind energy with increasing market penetration, which is consistent with Hirth (2016) and Schöniger and Morawetz (2022). Both effects are meaningful in size. In fact, our model predicts that sufficient reservoir hydro capacity allows to completely offset the negative effect of wind market penetration on wind value (see Figure 6). As expected, a stronger correlation of hourly electricity consumption and wind generation increases wind value factors and mitigates the negative effect of wind penetration on wind value. On the other hand, a more volatile or "peaky" generation pattern of wind is associated with lower wind value factors and a steeper drop of value factors at increasing market penetration. We find a negative effect of the clean gas-coal price ratio on wind value factors, but the aggregated nature of our proxy does not allow to attribute this to carbon or fuel prices. We find a small and negative effect of neighboring solar penetration on domestic wind value. Figure 6 visualizes the interaction effects of domestic wind/solar penetration with the control variables. Complete model results are presented in Table 3.

## 4.2 Solar

Confirming previous results, wind find that the value drop of solar is more pronounced than that of wind energy: A one p.p. increase in domestic solar market penetration over time causes a 1.39 p.p. drop in solar value factors. The *between* effect of average zone-level solar penetration is somewhat larger than the *within* effect: An average zone-level solar penetration of one p.p. above the sample average of 4% is associated with a 1.84 p.p. lower value factor. [6]

Solar cross-border spillovers are much larger than those of wind. For a simultaneous one p.p. increase in solar penetration across the border, domestic solar value factor is reduced by a remarkable 2.38 p.p. The *between* effect is very similar at 2.26 p.p. This pronounced spatial effect of solar is likely due to the strong correlation of solar generation across European bidding zones. Price-depressing solar electricity imports therefore more often add to the price pressure from domestic solar generation than wind does. Another reason for the larger spatial effect of solar could be that wind generation is associated with transmission congestion, while solar generation is not (Titz et al., 2024). This is not perfectly captured by the (time-invariant) interconnector capacity weights of neighboring generation described above and potentially reduces the cross-border effects of wind generation.

Like for wind, we find that higher interconnector capacity mitigates the negative effect of domestic solar penetration on solar value but exacerbates the negative effect of cross-border solar penetration on domestic solar value. That said, we also see wider confidence intervals in the right panel of Figure 4 as standard errors are potentially inflated by the near-collinearity of domestic and neighboring solar penetration. Unlike for wind, however, the trade-off between the negative effects of domestic and the negative effects of neighboring solar penetration on domestic solar value is almost perfect as the slopes of the two interaction effects shown in the right panel of Figure 4 and in Table 3 are almost identical. Similarly, the combined point estimates of the zone-level conditional effects shown in the bottom panel of Figure 5 are around -3.8 p.p. both for badly and well-interconnected zones. This is likely due to the strong correlation of solar generation profiles across space. In other words, well-interconnected zones are essentially subject to the same “peaky” solar generation profile than badly interconnected zones. Higher interconnector capacity thus appears to benefit wind but not solar value, which confirms the numerical modelling results of Hirth (2015).

Among our control variables, we can identify less drivers of the solar value drop compared to the wind model. We find a significant and positive lower-level interaction of solar penetration and hydro pumped storage capacity, which potentially reflects the positive effects of the recent commissioning of large pumped storage plants in Portugal and Switzerland on solar value. A higher solar variability – due to stronger diurnal and seasonal patterns at higher latitudes – has a negative effect of on value factor levels and exacerbates the negative effect of solar penetration on solar value. We find no significant effect of domestic wind penetration on solar value but a positive, albeit small effect of neighboring wind penetration.

---

[6] Comparing *within* and *between* effects of market penetration is not straightforward, however, as we only compute cross-level (i.e., *within***between*) and no higher-level (i.e., *between***between*) interactions. *Within* coefficients of market penetration are therefore effects at mean levels of interacted variables, while *between* coefficients do not control for interaction effects. Since *within* and *between* variables are uncorrelated by construction, potentially biased *between* coefficients do not affect our *within* coefficients.

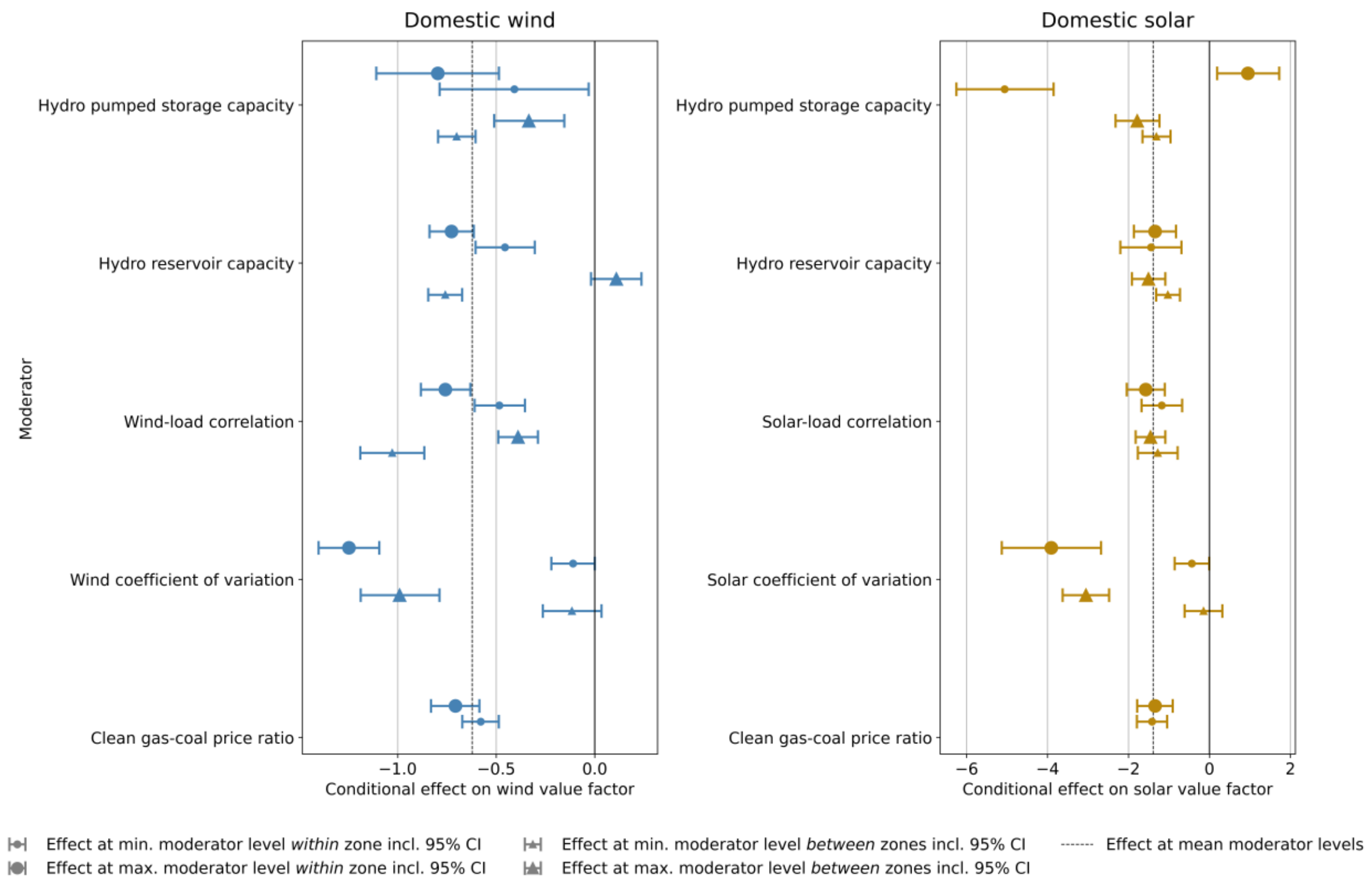

*Figure 6: Effects of domestic wind/solar penetration on domestic wind/solar value factors, conditional on moderator levels*

| | Wind value factor | Solar value factor |
|---|---|---|
| WITHIN EFFECTS | | |
| Domestic wind | -0.622*** (0.059) | -0.05 (0.046) |
| Neighboring wind | -0.486*** (0.042) | 0.162*** (0.055) |
| Domestic solar | 0.047 (0.06) | -1.392*** (0.22) |
| Neighboring solar | -0.151* (0.079) | -2.377*** (0.227) |
| Hydro pumped storage capacity | 0.023 (0.047) | 0.222*** (0.049) |
| Hydro reservoir capacity | 0.04 (0.046) | 0.028 (0.115) |
| Clean gas-coal price ratio | -0.008** (0.004) | 0.014** (0.007) |
| Wind-load correlation | 0.106*** (0.01) | |
| Wind coefficient of variation | -0.204*** (0.014) | |
| Domestic wind*Hydro pumped storage capacity | -0.93 (0.756) | |
| Domestic wind*Hydro reservoir capacity | -0.357 (0.241) | |
| Domestic wind*Clean gas-coal price ratio | -0.076 (0.061) | |
| Domestic wind*Wind-load correlation | -0.246** (0.122) | |
| Domestic wind*Wind coefficient of variation | -1.162*** (0.164) | |
| Solar-load correlation | | 0.347*** (0.024) |
| Solar coefficient of variation | | 0.02 (0.016) |
| Domestic solar*Hydro pumped storage capacity | | 12.533*** (2.183) |
| Domestic solar*Hydro reservoir capacity | | 0.755 (4.023) |
| Domestic solar*Clean gas-coal price ratio | | 0.043 (0.161) |
| Domestic solar*Solar-load correlation | | -0.467 (0.542) |
| Domestic solar*Solar coefficient of variation | | -1.106*** (0.306) |
| BETWEEN EFFECTS | | |
| Domestic wind (zone average) | -0.204 (0.179) | 0.246 (0.276) |
| Neighboring wind (zone average) | -0.29* (0.174) | 0.175 (0.299) |
| Domestic solar (zone average) | -0.23 (0.461) | -1.837** (0.785) |
| Neighboring solar (zone average) | 0.547 (0.56) | -2.256* (1.233) |
| Hydro pumped storage capacity (zone average) | 0.039 (0.059) | 0.05 (0.101) |
| Hydro reservoir capacity (zone average) | 0.024 (0.016) | -0.015 (0.083) |
| Wind-load correlation (zone average) | -0.034 (0.171) | |
| Wind coefficient of variation (zone average) | -0.062 (0.105) | |
| Solar-load correlation (zone average) | | -0.112 (0.229) |
| Solar coefficient of variation (zone average) | | 0.097 (0.236) |
| Interconnector capacity (zone average) | 0.002 (0.022) | -0.034 (0.043) |
| CROSS-LEVEL INTERACTIONS | | |
| Domestic wind*Hydro pumped storage capacity (zone average) | 0.545** (0.223) | |
| Domestic wind*Hydro reservoir capacity (zone average) | 0.241*** (0.035) | |
| Domestic wind*Wind-load correlation (zone average) | 1.766*** (0.473) | |
| Domestic wind*Wind coefficient of variation (zone average) | -1.792*** (0.356) | |
| Domestic wind*Interconnector capacity (zone average) | 0.19*** (0.04) | |
| Neighboring wind*Interconnector capacity (zone average) | -0.133*** (0.046) | |
| Domestic solar*Hydro pumped storage capacity (zone average) | | -0.558 (0.478) |
| Domestic solar*Hydro reservoir capacity (zone average) | | -0.649 (0.4) |
| Domestic solar*Solar-load correlation (zone average) | | -0.436 (0.654) |
| Domestic solar*Solar coefficient of variation (zone average) | | -4.051*** (0.749) |
| Domestic solar*Interconnector capacity (between zones) | | 0.412** (0.183) |
| Neighboring solar*Interconnector capacity (zone average) | | -0.415* (0.232) |
| Intercept | Yes | Yes |
| Adjusted $R^2$ | 0.461 | 0.589 |
| Observations | 2758 | 1941 |

*Table 3: Regression results. REWB regression model for wind and solar. Newey-West HAC standard errors in parentheses. Significance levels reported as *** for p < 1%, ** for p < 5%, and * for p < 10%.*

### 4.3 Exploring the sensitivity of results

In this section, we explore how sensitive our results are to alternative variable definitions and sample composition. The sensitivities that we consider relevant are the definition of spatial weights, the choice of aggregation period for wind and solar value factors, as well as the composition of our sample of European bidding zones.

#### 4.3.1 Sensitivity to spatial weighting

The spatial weights matrix defines between which panel units spatial effects are modelled and is hence crucial for our study. In our baseline models, we build on the binary spatial weights matrix, which is the approach used in most studies dealing with geographical units (Halleck Vega & Elhorst, 2015). Only relationships between direct geographical neighbors are modelled when a binary matrix is used, with each neighboring panel unit usually being weighted equally (Anselin et al., 2008). We use a weighted average of wind/solar penetration across direct neighbors instead, with weights given by bilateral interconnector capacity. In this sensitivity, we use an unweighted average to compare our baseline results against the most common definition of spatial weights in the literature.

We find a higher effect of neighboring wind penetration (-0.59 p.p. value drop per p.p. penetration increase vs. -0.48 p.p. in our baseline model) and a slightly smaller effect of domestic wind penetration (-0.58 vs. -0.62) on wind value. For solar, we find a smaller effect of neighboring solar penetration on solar value (-1.86 instead of -2.38) and a stronger effect of domestic solar (-1.71 vs. -1.39). The change in effects is as expected, since connectedness is positively correlated with neighboring wind levels and negatively correlated with neighboring solar levels in our sample (i.e., well-interconnected zones tend to have wind-rich neighbors and solar-poor neighbors). Using an unweighted average hence overestimates the effects of neighboring wind and underestimates the effect of neighboring solar.

#### 4.3.2 Leave-one-out analysis

In our large T, small N panel, the choice of bidding zones (i.e., N) to include in the panel can have a substantial impact on results – particularly when zones exhibit outlier values in variables of interest. We therefore run a leave-one-out sensitivity analysis in which we repeatedly estimate our model, each time omitting a different bidding zone from the sample. Figure A-2 shows the coefficients of interest (domestic and neighboring wind/solar penetration as well as the interaction effects of domestic and neighboring wind/solar penetration with interconnector capacity) for each of these model runs. The effects are similar in sign and size for our wind model, with the notable exception of DK_1, whose omission results in a stronger value drop of wind. This could be due to the stabilizing effect of Norwegian hydro storage on Danish wind value found by Greene and Vasilakos (2012), which we did not model. The coefficients in our solar model are somewhat more sensitive to the omission of single zones due to the smaller sample but all within the confidence intervals of the main effect when all zones are included.

#### 4.3.3 Sensitivity to aggregation period

As discussed above, we expected an annual instead of a monthly period of aggregation for wind and solar value factors to yield a more uncertain estimate of the wind and solar value drop (because of fewer observations), while a daily resolution should result in a smaller wind value drop (because wind generation affects daily base prices more strongly). Figure A-3 confirms these expectations: For wind and solar models, using an annual resolution gives more uncertain estimates of the effects of domestic and neighboring penetration as the number of observations drops to 230 and 162, respectively. A daily resolution, on the other hand, results in substantially smaller estimates of the wind value drop: Wind usually generates during most hours of the day and therefore reduces both its own revenue and average daily prices due to the merit-order effect, resulting in a small effect on the daily wind value factor. Using an hourly resolution would result in an effect of zero, since both numerator (market value) and denominator (average prices) of the value factor would be equally affected.

### *4.4 Robustness check: fixed effects estimation*

In this section we explore the robustness of results to fixed effects instead of random effects between-within estimation. Entity FE models are not biased by unobserved time-invariant entity heterogeneity but at the same time less efficient than our (partially pooled) REWB model, as between-entity variation is completely discarded in a FE model. To gauge the bias-efficiency trade-off we estimate our model controlling for entity fixed effects. Time-invariant variables, like interconnector capacity, cannot be estimated in this case as they are absorbed by zone dummies (or, equivalently, by within transformation). However, it is still possible to include interactions between time-varying and time-invariant variables, as done by Schöniger and Morawetz (2022). Interpretation of these interactions is feasible too, as the time-invariant base effects would be absorbed by the entity fixed effects and therefore do not need to be included separately. Table A-1 gives the results for FE model estimation and shows that both *within* effects and cross-level interactions are virtually identical to the REWB model, which is expected. *Between* effects, however, cannot be estimated.

# 5 Conclusion

In this study, we have estimated the effects that increasing wind and solar market penetration have had on wind and solar market value across 30 European bidding zones during 2015-2023. Besides the well-established negative effects of domestic market penetration, we find pronounced cross-border effects on wind and solar value. The negative effects of domestic and neighboring solar penetration on solar value are stronger than those of wind, as solar generation is more strongly correlated across time and space.

We have then investigated whether different capacities of bidding zones to import/export electricity alter the effects of market penetration on market value. In fact, we find that higher connectedness of bidding zones implies a trade-off: It mitigates the negative effect of domestic market penetration but exacerbates the effects of neighboring market penetration. The trade-off is almost perfect for solar as poorly and well-interconnected bidding zones see virtually the same aggregate effect of solar penetration on solar value because of the strong correlation of domestic and neighboring solar generation profiles. Well-interconnected zones see a smaller aggregate effect of wind penetration on wind value, however, as domestic and neighboring wind patterns are not as strongly correlated. Our findings hence imply moderate gains from higher connectedness only for wind generators.

In addition, we identify other factors that mitigate the value drop of wind and solar energy. Flexible pumped storage and reservoir hydro power, simultaneity of electricity consumption and renewable generation, as well as a smoother renewable generation profile all help to explain why some European electricity markets manage to integrate wind and solar energy better.

Not all factors shaping cross-border effects on renewable market value have been modelled in this study. Further research should continue to explore the spatial dimension by considering, for example, the spatial effects of flexibility and inflexibility on market value across interconnected electricity markets.

Our results have important implications for European energy policy, which strives for a low-carbon and integrated electricity market. While the benefits of market integration for consumers are more straightforward, they are less certain for renewable energy generators. Wind value is stabilized by more interconnection, but solar needs other, complementary efforts such as short-term energy storage to stabilize its value.

# Appendix

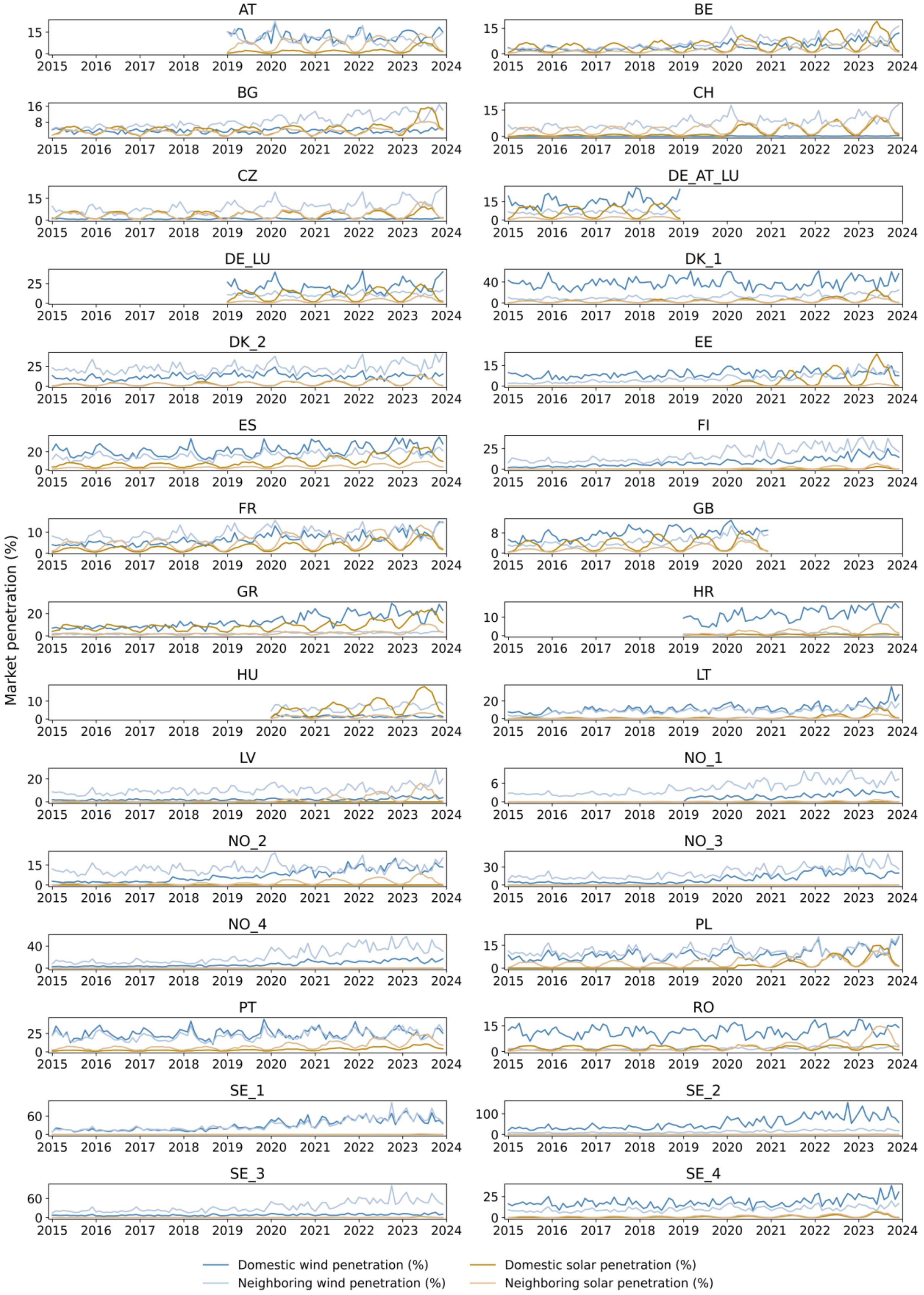

*Figure A-1: Time series of domestic and neighboring wind and solar penetration for European bidding zones*

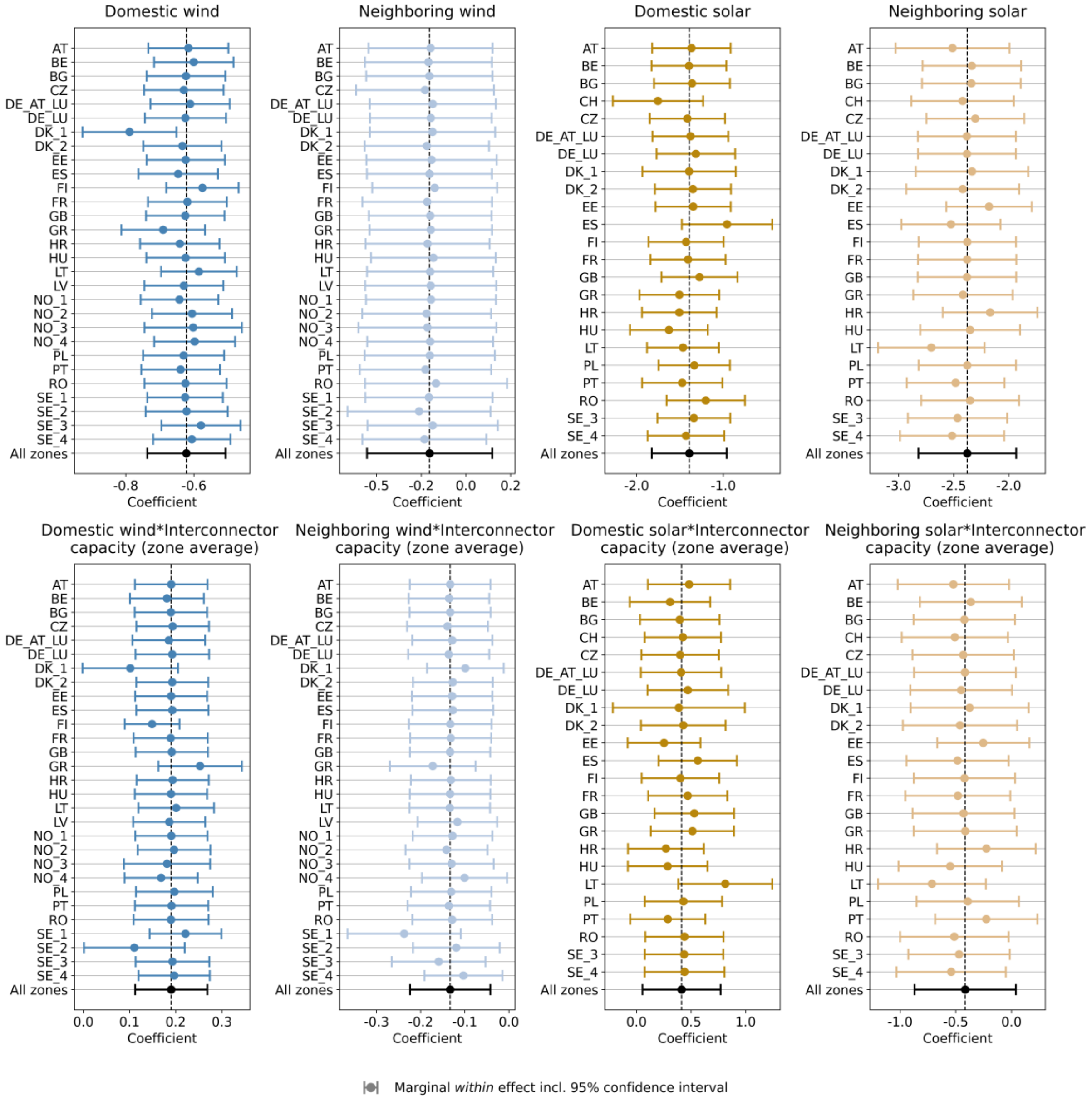

*Figure A-2: Leave-one-out-analysis for solar and wind models. Coefficients of interest when given zone is omitted from sample.*

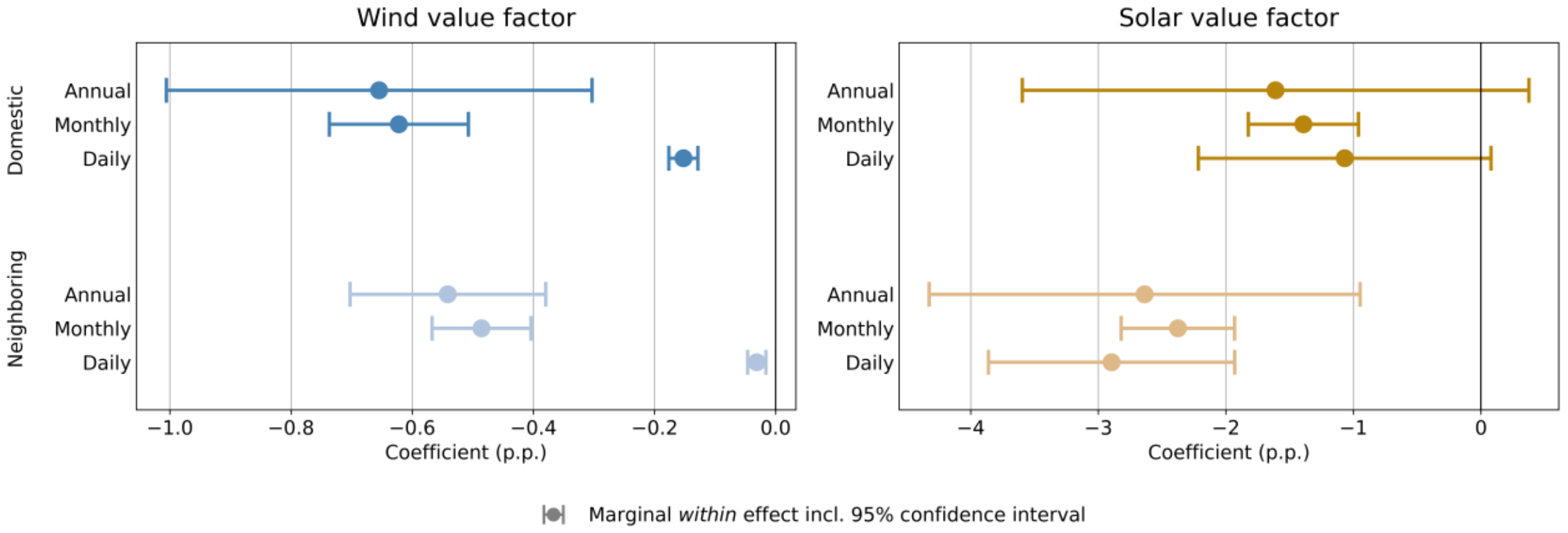

*Figure A-3: Marginal effects of domestic and neighboring wind/solar penetration at different temporal resolution*

| | Wind value factor | Solar value factor |
|---|---|---|
| WITHIN EFFECTS | | |
| Domestic wind | -0.622*** (0.058) | -0.052 (0.046) |
| Neighboring wind | -0.486*** (0.041) | 0.16*** (0.056) |
| Domestic solar | 0.047 (0.059) | -1.4*** (0.225) |
| Neighboring solar | -0.151* (0.078) | -2.374*** (0.227) |
| Hydro pumped storage capacity | 0.023 (0.048) | 0.223*** (0.049) |
| Hydro reservoir capacity | 0.04 (0.045) | 0.028 (0.115) |
| Clean gas-coal price ratio | -0.008* (0.004) | 0.014** (0.007) |
| Wind-load correlation | 0.106*** (0.01) | |
| Wind coefficient of variation | -0.204*** (0.014) | |
| Domestic wind*Hydro pumped storage capacity | -0.929 (0.79) | |
| Domestic wind*Hydro reservoir capacity | -0.357 (0.239) | |
| Domestic wind*Clean gas-coal price ratio | -0.076 (0.061) | |
| Domestic wind*Wind-load correlation | -0.246** (0.123) | |
| Domestic wind*Wind coefficient of variation | -1.162*** (0.164) | |
| Solar-load correlation | | 0.347*** (0.024) |
| Solar coefficient of variation | | 0.019 (0.016) |
| Domestic solar*Hydro pumped storage capacity | | 12.557*** (2.201) |
| Domestic solar*Hydro reservoir capacity | | 0.725 (4.024) |
| Domestic solar*Clean gas-coal price ratio | | 0.044 (0.161) |
| Domestic solar*Solar-load correlation | | -0.469 (0.534) |
| Domestic solar*Solar coefficient of variation | | -1.12*** (0.306) |
| CROSS-LEVEL INTERACTIONS | | |
| Domestic wind*Hydro pumped storage capacity (zone average) | 0.545** (0.221) | |
| Domestic wind*Hydro reservoir capacity (zone average) | 0.241*** (0.034) | |
| Domestic wind*Wind-load correlation (zone average) | 1.766*** (0.469) | |
| Domestic wind*Wind coefficient of variation (zone average) | -1.792*** (0.358) | |
| Domestic wind*Interconnector capacity (zone average) | 0.19*** (0.039) | |
| Neighboring wind*Interconnector capacity (zone average) | -0.133*** (0.046) | |
| Domestic solar*Hydro pumped storage capacity (zone average) | | -0.553 (0.491) |
| Domestic solar*Hydro reservoir capacity (zone average) | | -0.655 (0.405) |
| Domestic solar*Solar-load correlation (zone average) | | -0.427 (0.656) |
| Domestic solar*Solar coefficient of variation (zone average) | | -4.074*** (0.768) |
| Domestic solar*Interconnector capacity (between zones) | | 0.408** (0.183) |
| Neighboring solar*Interconnector capacity (zone average) | | -0.407* (0.234) |
| Intercept | Yes | Yes |
| Adjusted $R^2$ | 0.430 | 0.559 |
| Observations | 2758 | 1941 |

*Table A-1: Fixed effects regression results for wind and solar models. Newey-West HAC standard errors in parentheses. Significance levels reported as *** for p < 1%, ** for p < 5%, and * for p < 10 %.*

## Acknowledgements

This work was supported by the German Federal Ministry of Education and Research via the ARIADNE Project (FKZ 03SFK5K0-2).

## Author contributions

**Clemens Stiewe:** Conceptualization, Formal analysis, Visualization, Writing–original draft, Data curation, Project administration. **Alice Lixuan Xu:** Conceptualization, Formal analysis, Visualization, Writing–original draft, Data curation. **Anselm Eicke:** Conceptualization, Visualization, Writing–Review & editing. **Lion Hirth:** Conceptualization, Writing–Review & editing, Supervision.